\documentclass[journal=jpcbfk,manuscript=article]{achemso}
\usepackage[version=3]{mhchem} 
\usepackage{color}
\usepackage{lineno}
\setkeys{acs}{usetitle = true}
\usepackage[textsize=tiny,backgroundcolor=yellow]{todonotes}
\def\cm{cm$^{-1}$}
\def\ie{i.e.\ }

\newcommand*{\bra}[1]{\left<#1\right|}
\newcommand*{\ket}[1]{\left|#1\right>}
\newcommand*{\eref}[1]{eq~\plainref{#1}}
\newcommand*{\fref}[1]{Figure~\plainref{#1}}

\newcommand{\onlinecite}[1]{\hspace{-1 ex} \nocite{#1}\citenum{#1}} 
\newcommand*{\bref}[1]{ref~\onlinecite{#1}}
\author{J. Schulze}
\affiliation[Universit\"at Rostock]
{Institut f\"ur Physik, Universit\"at Rostock, D-18051 Rostock, Germany}
\author{O. K\"uhn}
\affiliation[Universit\"at Rostock]
{Institut f\"ur Physik, Universit\"at Rostock, D-18051 Rostock, Germany}
\email{oliver.kuehn@uni-rostock.de}

\title{Explicit Correlated Exciton-Vibrational Dynamics of the FMO Complex}
\keywords{Frenkel excitons, exciton-vibrational coupling, high-dimensional quantum dynamics, photosynthesis, FMO complex}
\begin{document}
\begin{abstract}

The coupled exciton-vibrational dynamics of a 3-site \textcolor{black}{Fenna-Matthews-Olson (FMO) model} is investigated using the numerically exact multilayer multiconfiguration time-dependent Hartree approach. Thereby the vibrational mode specific coupling to local electronic transitions is adapted from a discretized experimental spectral density. The solution of the resulting time-dependent Schr\"odinger equation including three electronic and 450 vibrational degrees of freedom is analyzed in terms of excitonic populations and coherences. Emphasis is put onto the role of specific ranges of vibrational frequencies. It is observed that modes between 160 and 300 \cm{} are responsible for the subpicosecond population and coherence decay. \textcolor{black}{Further, it is found that a mean-field approach with respect to the vibrational degrees of freedom is not applicable.}
\end{abstract}
%
\section{Introduction}
The discovery of long-lasting coherent oscillations in two-dimensional spectra obtained for a number of photosynthetic antenna complexes up to physiological temperatures \cite{engel07_782,panitchayangkoon11_20908,collini09:369} has triggered a host of theoretical investigations (for a review, see \bref{schroeter15_1}). A controversial discussion focussed on the role of vibrational degrees of freedom (DOFs). Traditionally, vibrations have been considered to form a heat bath, thus leading to phase and energy relaxation and  serving the purpose of directed downhill energy transfer.\cite{grondelle94,renger01_137} With the decoherence time of the observed oscillations considerably exceeding the expectations based on typical line  widths of electronic transitions, the coupling to vibrations was reconsidered with the result that it might be more important than anticipated based on the smallness of the Huang-Rhys (HR) factor.\cite{christensson12_7449,tiwari12_1203} In fact a special role was ascribed to a vibrational mode having a frequency that matches the electronic energy gap between neighbouring bacteriochlorophyll sites.\cite{christensson12_7449,chin13_113,chenu13_2019} Such a mode, which for the relevant sites one to three in the FMO complex should have a frequency of about 180 \cm, could facilitate vibrationally-assisted resonance transfer. In terms of the eigenstates of the Coulomb-coupled excitonic and vibronic excitations there might be a considerable quantum state mixing \cite{polyutov12_21,schulze14_045010} making an analysis of the observed spectra in terms of pure electronic or vibronic excitations difficult.\cite{butkus14_034306,tempelaar14_12560} Interestingly, a popular  experimentally determined spectral density features a peak in the 180 \cm{} region.~\cite{wendling00_5825}

Simulations of coupled exciton-vibrational dynamics are usually performed using  non-pertur\-ba\-ti\-ve and non-Markovian methods such as the hierarchy equation of motion \cite{kreisbeck12_2828,struempfer12_2808,Ishizaki:2010fx} or path integral approaches.\cite{thorwart09:234} Here, the dynamics is obtained explicitly only for the relevant system, whereas the bath is traced out in the reduced density operator.\cite{may11} Apart from a few examples, where a single vibrational mode has been kept alive in the relevant system\cite{kuhn96_99,butkus14_034306,nalbach15_022706}, the latter consists of the excitonic DOFs only. Hence, there is no access to the explicit dynamics of the vibrational modes within these system-bath approaches. In passing we note that the inclusion of a single vibrational mode into the relevant system has, of course, some arbitrariness with respect to the choise of the modes' properties.

In the present contribution we study the coupled exciton-vibrational dynamics from a completely different perspective. Specifically, we solve the time-dependent Schr\"odinger equation for a high-dimensional model of the FMO complex. This allows us, for the first time, to access information on the dynamics of electronic ground/excited  state vibrational/vibronic excitations during excitation energy  transfer in this complex. This becomes possible due to the special structure of the Frenkel exciton Hamiltonian, which is ideally suited for the combination with the multilayer multiconfiguration time-dependent Hartree (ML-MCTDH) wave packet propagation method.\cite{meyer90_73,beck00_1,wang03:1289,manthe08_164116}

\section{Theory}
We will use the standard  Frenkel exciton Hamiltonian to describe an aggregate with $ N_{\rm agg} $ sites (site index $m$), each site having the excitation energy $E_m$, and different sites being coupled by the Coulomb interaction $J_{mn}$:
\begin{equation} 
\label{eq:exham}
H_{\rm ex} =\sum_{m,n=1}^{N_{\rm agg}}(\delta_{mn} E_m + J_{mn})\ket{m}\bra{n}  \,.
\end{equation}
Local electronic states are restricted to the ground, $\ket{g_m}$, and excited,  $\ket{e_m}$, states,  i.e.\ the local or diabatic one-exciton states are given by $\ket{m}=\ket{e_m}\prod_{n\ne m}\ket{g_n}$. For the present simulations we will adopt the 8-site FMO Hamiltonian reported by Moix and coworkers.\cite{moix11_3045} It is reduced to a 3-site model (sites 1-3) as suggested by the simulations of population dynamics in \bref{moix11_3045} \textcolor{black}{(for full set of parameters see Supplementary Information (SI)).  Indeed, starting from an initial population of site one only, the populations of individual sites other than 1-3 never exceed about 12\% within the first 1~ps. This finding is supported by the decomposition of eigenstates into the local basis for full and reduced models as given in Table S1 (SI). In the 3-site model  the exciton Hamiltonian matrix is given by (in \cm)}
 
\begin{eqnarray}
H_{\mathrm{ex}}=\left(  
    \begin{array}{ccc}
	{310} & {-98} & {6}\\
      {-98} &{230}& {30} \\
      {6}& {30} & {0}  
    \end{array}                    
  \right) \, .
\end{eqnarray}

Diagonalization of this  matrix  yields the one-exciton eigenstates,  $\ket{\alpha}$,  whose decomposition into the local states $\ket{m}$ is given by $\ket{\alpha} = \sum_{m} C_{\alpha,m}\ket{m}$ and shown together with the eigenenergies in \fref{fig2}. The local vibrations at site $m$ are described in harmonic approximation by the set of dimensionless normal mode coordinates $ \{ Q_{m,\xi} \} $ with frequencies $ \{ \omega_{m,\xi} \} $, \ie the vibrational Hamiltonian reads
\begin{equation}
\label{eq:hvib}
H_{\rm vib} =\sum_m \sum_{\xi \in m}\frac{\hbar\omega_{m,\xi}}{2} \left( - \frac{\partial^2}{\partial Q_{m,\xi}^2}+ Q_{m,\xi}^2\right) \, .
\end{equation}
Here, the notation $\xi \in m$ indicates the summation over all vibrational modes of site $m$.
Exciton-vibrational coupling is accounted for within the linearly shifted oscillator (Huang-Rhys) model given by
\begin{equation}
\label{eq:hexvib}
H_{\rm ex-vib} = \sum_m \sum_{\xi \in m} \hbar \omega_{m,\xi} \sqrt{2 S_{m,\xi}}\,Q_{m,\xi} \ket{m}\bra{m}\, .
\end{equation}
 The coupling of a particular mode to the local electronic transition and thus the degree of vibronic excitation, is characterized by the HR factor $S_{m,\xi}$. In system-bath models this coupling is typically expressed via the spectral density\cite{may11}
\begin{equation}
\label{eq:jomega}
	J_m(\omega) = A \sum_{\xi\in m}   S_{m,\xi} \delta(\omega-\omega_{m,\xi})\, ,
\end{equation}
where $A$ is a constant to be specified below.
Note that $J_m(\omega)$ corresponds to the spectral density for monomeric bacteriochlorophyll $a$ in the actual FMO protein environment. There are several calculations of this spectral density based on classical molecular dynamics and  taking into account the protein and solvent environment.~\cite{olbrich11_1771,renger12_14565,rivera13_5510,valleau12_224103} However, the  reported results differ considerably, with respect to their structure and integrated intensity; the latter gives the total HR factor for site m: $S_{\rm tot}=A^{-1}\int d\omega J_{m}(\omega) = \sum_{\xi \in m} S_{m,\xi}$.~\cite{may11} Therefore, we will use the experimentally determined spectral density of Wendling et al. \cite{wendling00_5825} shown in \fref{fig2}. It has been obtained from low-temperature site-selected fluorescence, measured for the energetically lowest pigment of the complex under the assumption that the Coulomb coupling to the other pigments can be neglected. The total HR factor was determined as 0.42. Note, however, that it has recently been pointed out that the correction to the HR factor of particular modes due to the Coulomb coupling can be substantial, i.e. for modes in the 180 \cm{} region up to a factor of 1.5.\cite{schulze14_045010} Notice further that the experimental spectral density covers the range up to about 350 \cm{} only, whereas some calculations predict distinct structure up to 2000 \cm.~\cite{olbrich11_1771} However, excitonic transition energies between strongly coupled pigments are essentially located in the range up to 300 \cm, what justifies to neglect higher frequency modes. \textcolor{black}{In \bref{valleau12_224103} a detailed comparison of this low-frequency part of $J(\omega)$ with results of Quantum Mechanics/Molecular Mechanics calculations was given. Although different in details, the calculations could in particular be used to justify the assumption of the HR model leading to \eref{eq:hvib} and \eref{eq:hexvib}.} In the present model  we will use the original spectral density from \bref{wendling00_5825} and discretize it into 150 modes within the interval $[2:300]$ \cm. The amplitudes of the individual HR factors have been adjusted homogeneously via the constant $A$ in \eref{eq:jomega} such as to  preserve $S_{\rm tot}=0.42$ upon summation. 

The time-dependent Schr\"odinger equation will be solved employing the ML-MCTDH method (for a review, see \bref{meyer11_351}). Here, the  the state vector is expanded in terms of the local exciton basis according to
\begin{eqnarray}
	|\Psi({\bf Q};t) \rangle=\sum_{m} \chi_{m}({\bf Q};t) \, |m\rangle	\, ,
\end{eqnarray}
where the  nuclear coordinates are comprised into the $D=N_{\rm agg}\times N_{\rm vib}=3 \times 150 = 450$ dimensional vector $\mathbf{ Q}$.  The nuclear wave function for each diabatic state is  expanded into MCTDH form 
\begin{equation}
\label{eq:psiMCTDH}
\chi_\alpha(\mathbf{ Q},t) = \sum_{j_1 \ldots j_D}^{{n_{j_1} \ldots n_{j_D}}}
C^{(\alpha)}_{j_1,\ldots,j_D}(t) \phi^{(\alpha)}_{j_1}(Q_1;t) \ldots \phi^{(\alpha)}_{j_D}(Q_{D};t) \, .
\end{equation}
Here, the $C^{(\alpha)}_{j_1,\ldots,j_D}(t)$ are the time-dependent expansion coefficients weighting the contributions of the different Hartree products, which are composed of $n_{j_{k}}$ single particle functions (SPFs), $\phi^{(\alpha)}_{j_k}(Q_k;t)$, for the $k$th degree of freedom in state $\alpha$. In ML-MCTDH the SPFs itself describe multi-dimensional coordinates that are expanded into MCTDH form.~\cite{wang03:1289,manthe08_164116,vendrell11_044135}
This nested set of expansions can be represented by so-called ML-MCTDH trees.~\cite{manthe08_164116}  The particular choice of this tree may have a strong influence on the required numerical effort.~\cite{schroeter15_1} In the following simulations we use a grouping according to the magnitude of the HR factor as detailed in the SI. There we also give the other parameters of the ML-MCTDH setup. Wave packet propagations have been performed using the Heidelberg program package \cite{mctdh84}. Temperature effects due to the thermal population of vibrational states in the electronic ground state are not included. In all case the initial conditions has been a vertical Franck-Condon transition at site 1 and the propagation time was 1 ps. \textcolor{black}{Convergence of the ML-MCTDH setup has been monitored by means of the natural orbital populations \cite{beck00_1}. In the full/reduced model the largest  population of the least occupied natural orbital was $\sim 10^{-4}$/$\sim 10^{-6}$.}

\begin{figure}[hbt]
  \includegraphics[width=0.9\textwidth]{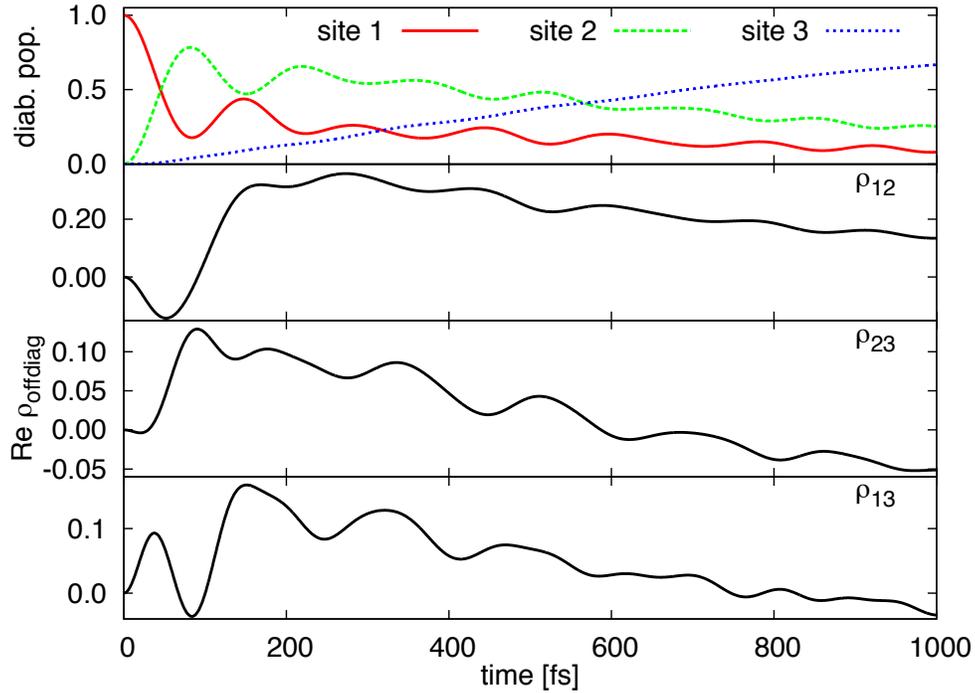}
  \caption{\label{fig1}
  Upper panel:  Diabatic population dynamics for the 3-site FMO model. Lower panels: Coherence dynamics as indicated in terms of the off-diagonal elements of the electronic density matrix in the diabatic, i.e.\ site-based, representation. Note that the long-time behaviour of the coherences reflects the mixing of the adiabatic states with respect to the diabatic ones (cf.\ \fref{fig2}).
  }
\end{figure}

\section{Results and Discussion}
\fref{fig1} shows the diabatic population and coherence dynamics of the 3-site model. The diabatic populations show a clear beating between sites 1 and 2, overlaid by a damping of both oscillation amplitudes. For site 3 one observes only rather small amplitude oscillations on top of an otherwise monotonously  increasing population. A Fourier decomposition of the population dynamics yields a dominant component at 220 \cm{} (period 151 fs). This does not agree with the bare electronic oscillation period, which would be 160 fs (208 \cm). It is this difference which actually reflects the modified energy level structure of the coupled exciton-vibrational system. In principle this oscillatory behaviour is qualitatively similar to previously published results.~\textcolor{black}{\cite{ishizaki09_17255,moix11_3045,dijkstra12_073027} There are quantitative differences due to different Hamiltonian parameters \cite{ishizaki09_17255,dijkstra12_073027} and spectral densities.~\cite{ishizaki09_17255,moix11_3045,dijkstra12_073027} In addition, the present model assumes a discretized bath and does not include temperature, which might be the reason for the population oscillations to last over the whole time interval.}

The coherences, $\rho_{mn}(t)=\langle m \ket{\Psi(t)}\bra{\Psi(t)} n \rangle $, between the different sites also show  damped oscillations as seen in the lower panels of 
\fref{fig1}. The oscillations of coherence density matrix elements have been related to the off-diagonal peaks in the two-dimensional spectroscopy of the FMO complex.\cite{panitchayangkoon11_20908,hein12_023018} Note, however, that the actual assignment to specific pairs of exciton eigenstates depends on the parametrisation of the Hamiltonian. Experimentally, in \bref{panitchayangkoon11_20908} oscillations at about 160 \cm{} have been reported, irrespective of temperature in the range between 77~K and 150~K.  In the present case, which formally corresponds to zero temperature, the dominant frequencies in this range are  at  220 \cm{} ($\rho_{12}$) and at 190 \cm{} ($\rho_{23}$); both frequencies occurring for $\rho_{13}$. We also observe that the amplitudes of these oscillations decay considerably within the given time window due to the Coulomb coupling induced mixing between electronic and vibronic/vibrational excitations. Apparently, the finite but large number of vibrational DOFs is sufficient to mimic excitonic energy relaxation and coherence dephasing.

\begin{figure}[tbh]
  \includegraphics[width=\textwidth]{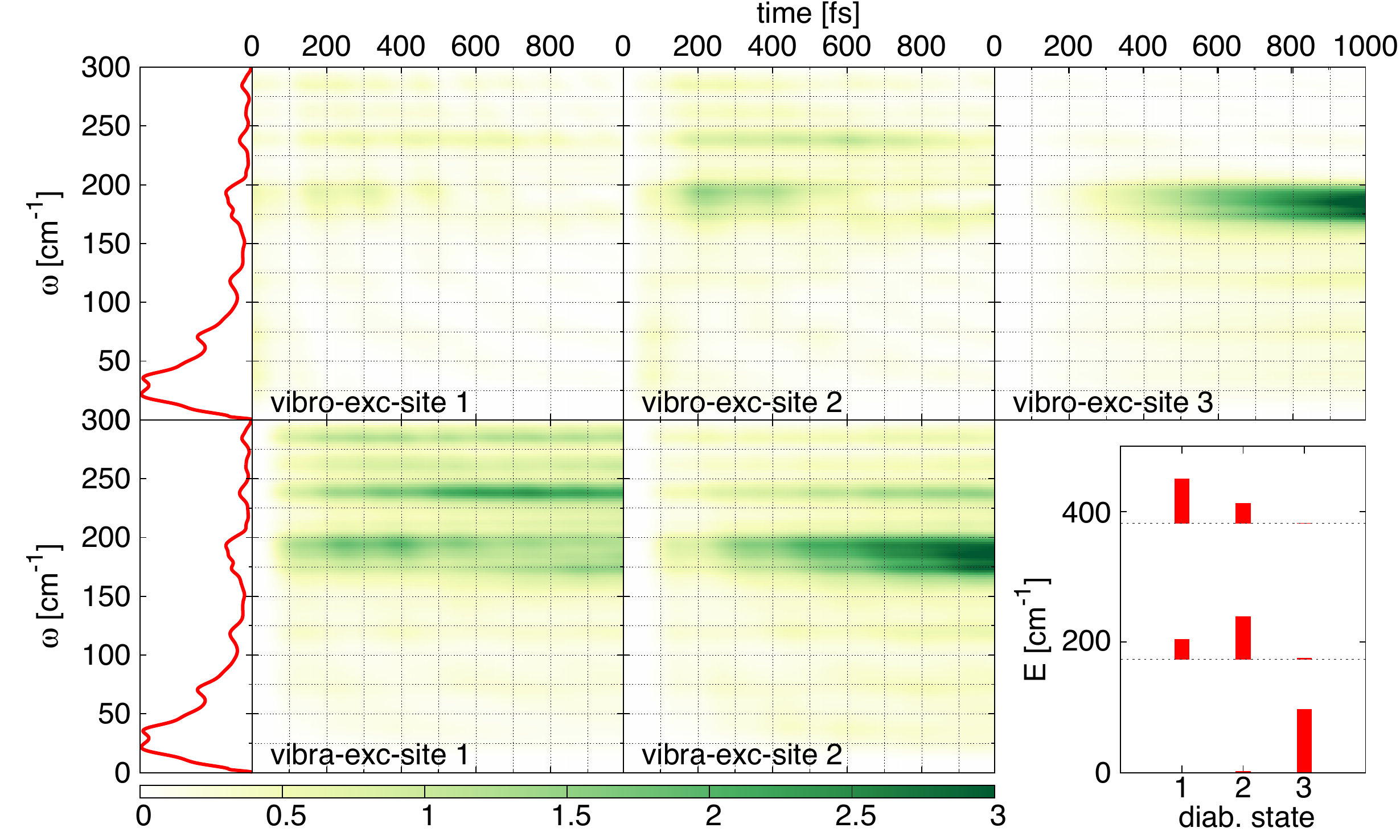}
  \caption{\label{fig2}
  Vibronic (upper row) and vibrational (lower row) excitation dynamics of the 3-site FMO model using a discretization of the spectral density into 150 equally spaced modes in the interval $[2:300]$ \cm{} for each monomer. Note that the vibrational excitation of site 3 is close to zero and not shown. The spectral density is given in the left panels. The lower right panel shows the spectrum
 of one-exciton eigenstates and their decomposition into amplitudes of (local) diabatic states. The color code covers the interval from 0 to 3 \cm{}; zero point energy has been subtracted.}
\end{figure}

In order to investigate the vibronic/vibrational excitation more closely we consider the dynamics in the different potential energy surfaces. Specifically, we define the operator 
\begin{equation}
\label{eq:vibra}
 H^{\rm (vibra)}_m=\sum_{\xi \in m}\frac{\omega_{m,\xi}}{2}\left(-\frac{\partial^2}{\partial Q_{m,\xi}^2}+Q_{m\xi}^2\right)
 (1-|m\rangle\langle m| )\, ,
\end{equation}
whose expectation value gives  the vibrational energy in the electronic ground states for monomer $m$, no matter the electronic population of the other monomers $n\ne m$. 

The vibronic energy at monomer $m$ follows from the expectation value of 

 \begin{equation}
 \label{eq:vibro}
 H^{\rm (vibro)}_m=
\sum_{\xi \in m} \frac{\hbar \omega_{m,\xi}}{2}\left(-\frac{\partial^2}{\partial Q_{m,\xi}^2}+Q_{m,\xi}^2+ 2 \sqrt{2 S_{m,\xi}}  Q_{m,\xi} \right) |m\rangle\langle m| \, .
 \end{equation}
 
\fref{fig2} shows the dynamics of vibronic and vibrational excitations for the situation of \fref{fig1}.  First,  we notice that the degree of excitation is rather small. Expectation values of the energy do not exceed 3 \cm{} for individual modes. Second, there is a marked difference between the individual sites and the dynamics in their respective potential energy surfaces. This can be explained either in a dynamical or an eigenstate picture. Within a dynamical picture vibrational excitation in the electronic ground state corresponding, e.g., to the site of initial electronic excitation, is a consequence of the finite transfer time between sites 1 and 2. If the latter is longer than the time scale of vibrational motion the initial vibronic wave packet moves out of the original Franck-Condon window. Thus, upon deexcitation during the transfer event, vibrationally excited states are populated in the electronic ground state.   We emphasise that the oscillations due to exciton transfer observed in \fref{fig1} are reflected as well in the vibrational and vibronic excitations. Eventually, the exciton population is accumulated at site 3 and so is the vibronic excitation. The trapping at site 3 comes along with negligible vibrational excitation at this site (not shown). However,  the nonequilibrium vibrational excitation at sites 1 and 2 is maintained even at 1 ps. Note that in order to describe vibrational relaxation, which would be relevant at longer time scales, one should include anharmonic interactions among the harmonic oscillator modes. 

Third,  considerable excitation is restricted to certain mode frequency ranges and here most notably between about 160 to 200 \cm. In the dynamical picture this could be explained by the fact that the electronic time scale is about 160 fs, i.e. only modes with higher frequencies have a chance to leave the Franck-Condon window and, therefore, become vibrationally excited. The selectivity of strong vibronic excitation in particular at site 3 is better explained in terms of a resonance effect, i.e. vibrationally-assisted  exciton transfer. This is similar to the vibronic enhancement of exciton transport discussed for allophycocyanine dimers in \bref{womick11_1347}. In  \fref{fig2} we also show the electronic eigenenergies and the decomposition of eigenstates into the local exciton basis. The energetic separation between adjacent eigenstates is around 200 \cm, which explains the preference of excitation for this region of the spectral density. 

\begin{figure}[hbt]
  \includegraphics[width=\textwidth]{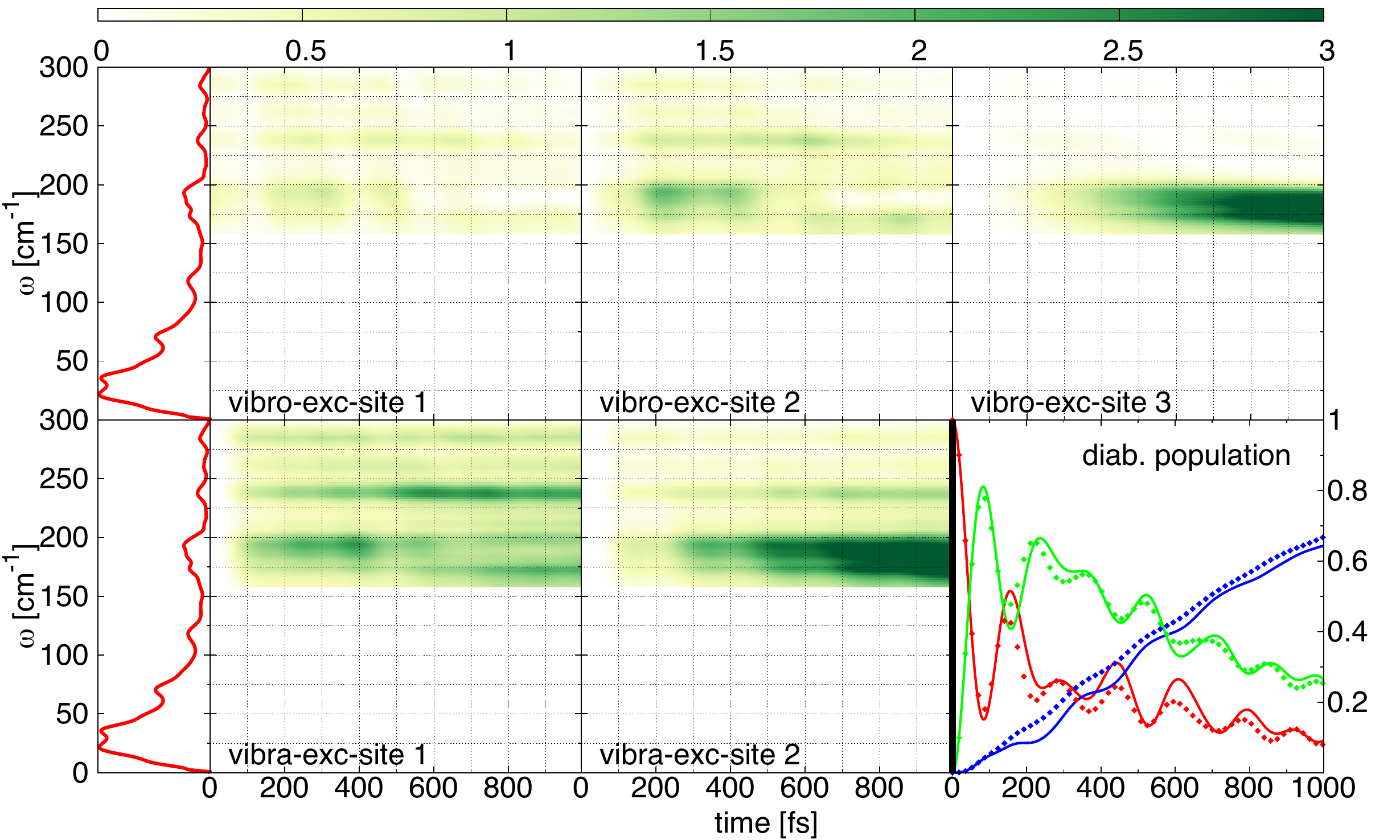}
  \caption{\label{fig3}
  Vibronic (upper row) and vibrational (lower row) excitation dynamics of the 3-site FMO model using a reduced description where only vibrational modes with frequencies in the interval $[159:300]$ \cm{} were included. The lower right panel compares the diabatic populations for full (dots) and reduced (lines) models. The color code covers the interval from 0 to 3 \cm{}; zero point energy has been subtracted.
  }
\end{figure}

In order to investigate the role of the high frequency part of the spectral density closer, we have studied a reduced model where only modes in the range $[159:300]$ \cm{} were included. The resulting dynamics is shown in \fref{fig3}. The lower right panel compares full and reduced models in terms of the diabatic population dynamics. Clearly, the differences between the two models are small and can be found mostly in the amplitudes of the oscillations, which are generally larger in the reduced model.

The vibrational and vibronic excitation also behaves similar, with the main difference being higher excitation energies obtained for the reduced model which reaches 4 \cm{} per mode. 
Comparing \fref{fig2} with \fref{fig3} this can be interpreted as the effect of vibrational energy redistribution, which is more pronounced in the full model due to the larger spectral range that is available. We emphasise again that for a single site the vibrational DOFs are decoupled by construction of the model. Vibrational energy redistribution is possible solely due to the Coulomb coupling between the sites.

\begin{figure}[hbt]
  \includegraphics[width=\textwidth]{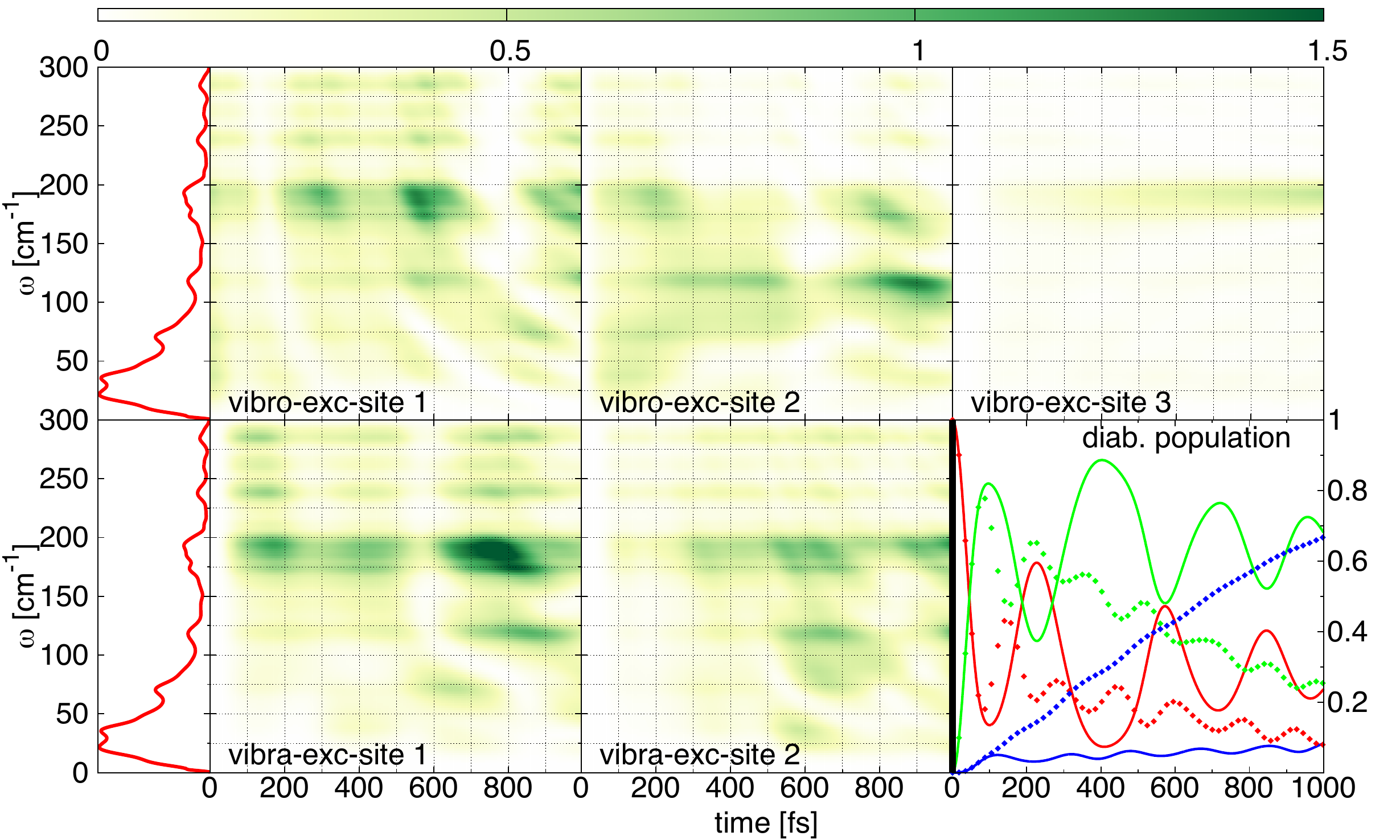}
  \caption{\label{fig4}
 \textcolor{black}{ Vibronic (upper row) and vibrational (lower row) excitation dynamics of the 3-site FMO model using the TDH approximation, \eref{eq:psiTDH}. The lower right panel compares the diabatic populations for the ML-MCTDH (dots) and TDH (lines) models. In contrast to \fref{fig2} and \fref{fig3} the color code covers the interval from 0 to 1.5 \cm{} only; zero point energy has been subtracted.}
  }
\end{figure}

\textcolor{black}{The ML-MCTDH approach provides a many-body wave function which is exact in the sense of the given numerical convergence thresholds. One might ask to what extent correlations actually matter. After all the isolated monomer is described by uncoupled vibronic excitations within the present model. The effect of Coulomb coupling on the correlations between vibrational DOFs is most straightforwardly accessed by comparison with the time-dependent Hartree (TDH) approximation to the wave function in \eref{eq:psiMCTDH} which reads}
\begin{equation}
\label{eq:psiTDH}
\chi_\alpha^{\rm (TDH)}(\mathbf{ Q},t) = 
C^{(\alpha)}(t) \phi^{(\alpha)}_1(Q_1;t) \ldots \phi^{(\alpha)}_D(Q_{D};t) \, .
\end{equation}
\textcolor{black}{
In passing we note  that the CPU time is reduced by a factor of about 600 when using \eref{eq:psiTDH} instead of \eref{eq:psiMCTDH}. However, the resulting dynamics shown in \fref{fig4} is not at all close to the  ML-MCTDH simulation in \fref{fig2}. We start by  comparing the population dynamics. Not only that the oscillation period for population exchange between sites 1 and 2 changes, the  exciton population also stays at sites 1 and 2 for a much longer time in TDH, i.e. the population trapping at site 3 is not very efficient within the first 1 ps. This can be attributed to lack of sufficient vibrational energy redistribution, which comes along with the less flexible TDH wave function. This is further supported by the vibrational and vibronic energy distributions shown in \fref{fig4}. First, the degree of excitation is much smaller than in \fref{fig2}; notice the different scales in the two figures. Second, apart from the oscillation pattern which is imprinted by the long-lasting oscillations of the populations, the spectral distribution of vibrational and vibronic excitations is also modified. Thus we conclude that the TDH approximation is not suitable for the description of the present FMO model.
}

\section{Conclusions}
In summary, we have presented a numerically exact quantum dynamics simulation of the Coulomb-coupled exciton-vibrational dynamics of a 3-site model of \textcolor{black}{the FMO complex at zero temperature.  This has become possible by combining the simplicity of the Frenkel exciton/linearly displaced harmonic oscillator Hamiltonian with the ML-MCTDH approach, thereby incorporating vibronic coupling via an experimentally determined spectral density. Adopting this model, numerically exact refers to the used convergence threshold for the natural orbital populations within ML-MCTDH.}

Most notably we have found that typical time scales for excitonic energy and phase relaxation can be reproduced by a finite discretization of an experimental spectral density. Further, in accordance with previous suggestions \cite{christensson12_7449, rey13_903} modes in the 180 \cm{} region play an important role for the transfer. The novel feature of explicit full-dimensional quantum dynamics used in the present simulation, however, is that it gives access to the dynamics of all DOFs. This allowed to discover that a large number of modes between 160 and 300 \cm{} are actually excited in the electronic ground and excited states. \textcolor{black}{This finding points to a possible dimensionality reduction in future studies, which may facilitate to account for all eight sites of the FMO. }In other words, the reduction to a single mode represents a rather strong approximation. 
\textcolor{black}{However, a simplification in terms of the mean-field description as provided by the TDH approximation is not suitable as it gives qualitatively different results. In other words, there is a strong correlation in the many-body dynamics of  excitonic and vibrational DOFs.}
We conclude by pointing out that the present simulation provides further evidence for the exceptional role that is played by vibrational and vibronic excitations in photosynthetic  energy transfer.~\cite{huelga13_181} 
\begin{acknowledgement}
The authors thank the Deutsche Forschungsgemeinschaft (DFG) for financial support through the Sfb 652. We gratefully acknowledge stimulating discussion with Marco Schr\"oter.
\end{acknowledgement}

\begin{suppinfo}
The appendix provides details of the exciton Hamiltonian and the ML-MCTDH computational setup.
\end{suppinfo}
%
\providecommand*\mcitethebibliography{\thebibliography}
\csname @ifundefined\endcsname{endmcitethebibliography}
  {\let\endmcitethebibliography\endthebibliography}{}

\clearpage\newpage
%
%
%
%
\section*{Table of contents entry}
\begin{figure}
\includegraphics[width=8.5cm]{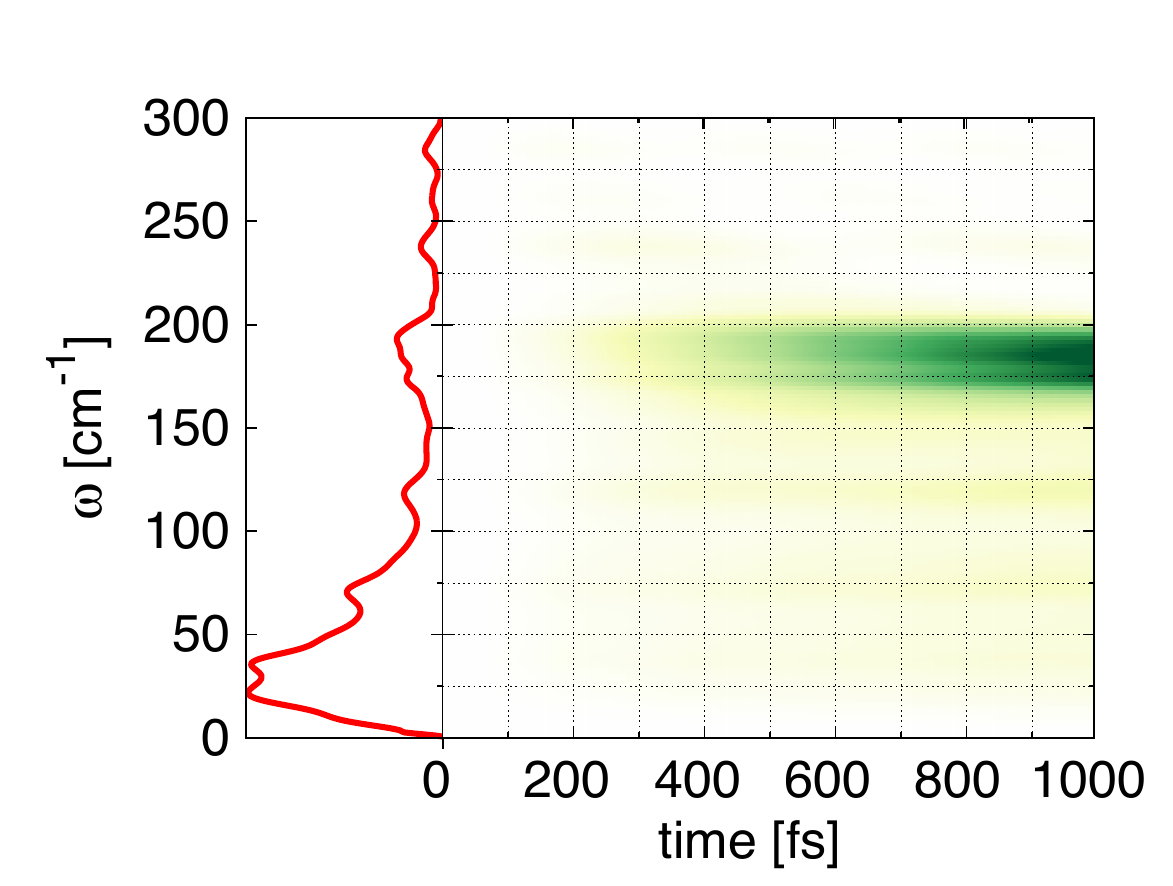}
\caption*{High-dimensional quantum dynamical simulations are performed for a FMO model which is based on an experimental spectral density (left panel). In the right panel the vibronic excitation is shown for  the terminal site of the complex.}
\end{figure}
\end{document}